# Adaptive Step Size Incremental Conductance Based Maximum Power Point Tracking (MPPT)

Eungkyun Kim
*Electrical and Computer Engineering Department*
*Tennessee Technological University*
Cookeville, USA
ekim43@students.tntech.edu

Morgan Warner
*Electrical and Computer Engineering Department*
*Tennessee Technological University*
Cookeville, USA
mawarner42@students.tntech.edu

Indranil Bhattacharya
*Electrical and Computer Engineering Department*
*Tennessee Technological University*
Cookeville, USA
ibhattacharya@tntech.edu

***Abstract*—Extracting maximum power available from photovoltaic arrays requires the system operating at the maximum power point (MPP). Therefore, finding the MPP is necessary for efficient operation of PV arrays. The MPP changes with multiple environmental factors, mainly temperature and irradiance. Traditionally, the incremental conductance technique with fixed step size was used to find the MPP, which suffers from a trade-off between speed of convergence and accuracy. In this work, we propose an incremental conductance maximum power point tracking (MPPT) algorithm with a variable step size, which adaptively changes the step size after each iteration based on how far away the current operating point is from a new MPP. This mitigates the aforementioned trade-off drastically by achieving faster convergence speed without the loss of accuracy. A series of simulations involving variations in temperature and irradiance were performed using MATLAB, and the speed of convergence and accuracy were compared with the traditional IC technique.**

***Keywords—incremental conductance, adaptive step size, convergence speed, maximum power point tracking, MPPT, photovoltaic array.***

## I. INTRODUCTION

A photovoltaic (PV) array is composed of several connected PV cells and converts sunlight directly into electricity [1], and the voltage and current at which the PV array operates are determined by the impedance of the load [2]. A typical current to voltage (I-V) characteristic curve of a PV array at standard test condition is shown in Fig. 1. In an open circuit condition, we obtain the maximum voltage, denoted by $V_{oc}$, and the maximum current is obtained in a short circuit condition, denoted by $I_{sc}$. The operating point of a PV array then lies between ($I_{sc}$, 0) and (0, $V_{oc}$). On the curve, there exists only one point where the product of voltage and current is maximum, and this particular point is called maximum power point (MPP). The voltage and current at the MPP are denoted by $V_{mp}$ and $I_{mp}$, respectively. Operating a PV array at any point other than the MPP would mean that the extracted power is not the maximum.

As shown in Fig. 2 and 3, the I-V curve of a PV array changes with varying environmental factors, such as temperature and irradiance, therefore the MPP of a PV array changes as well. Since any practical PV system would be under conditions in which these factors constantly change, the operating point of a PV array needs to be adjusted to the new MPP. The process to identify the new MPP is called maximum power point tracking (MPPT).

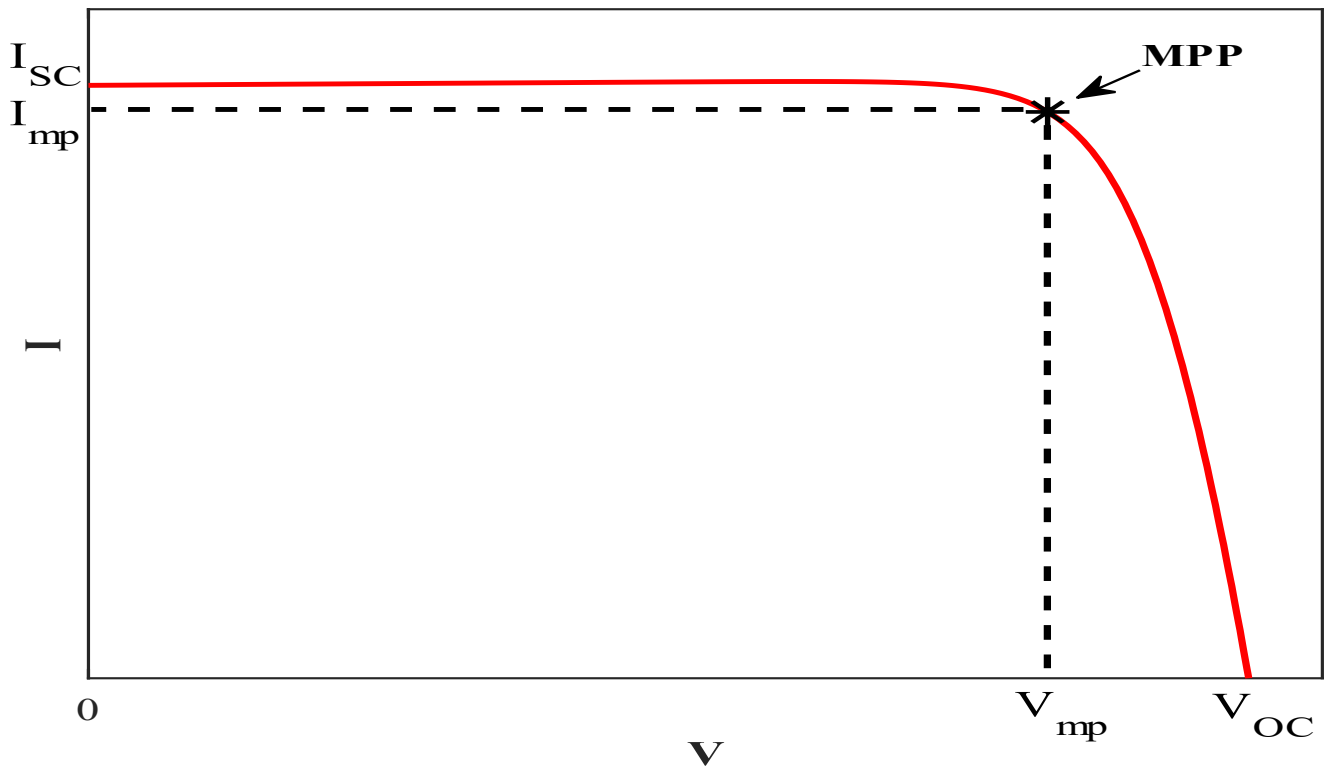

Fig. 1. I-V characteristic curve of a PV array

Many MPPT techniques have been proposed with differences in complexity, cost, convergence speed, and overall output efficiency [3]. For example, Perturb & Observe technique finds the MPP by observing the voltage level and perturbing it until it grasps the MPP [4]. While this method has been used extensively due to the ease of implementation, it suffers from sustained oscillation around the MPP and fast tracking versus oscillation trade-offs [5].

Another technique is incremental conductance (IC) whereby the MPP is found by comparing the incremental conductance with the instantaneous conductance. Because the derivative of power with respect to voltage ($\Delta P/\Delta V$) at the MPP must be zero [6], applying the product rule and chain rule, we can write the following equation:

$$\frac{dP}{dV} = \frac{d}{dV}(VI) = 1 + V\frac{dI}{dV} = 0 \tag{1}$$

Approximating dI/dV by choosing a small step size, $\Delta I/\Delta V \approx dI/dV$, yields:

$$\frac{\Delta I}{\Delta V} = -\frac{I}{V} \tag{2}$$

which is the condition for the MPP. This means that the operating point is at the MPP if the incremental conductance is equal to the instantaneous conductance. Similarly, if the incremental conductance is less than the instantaneous conductance, the operating point is to the left of the MPP, therefore the operating voltage needs to be increased, and vice versa.

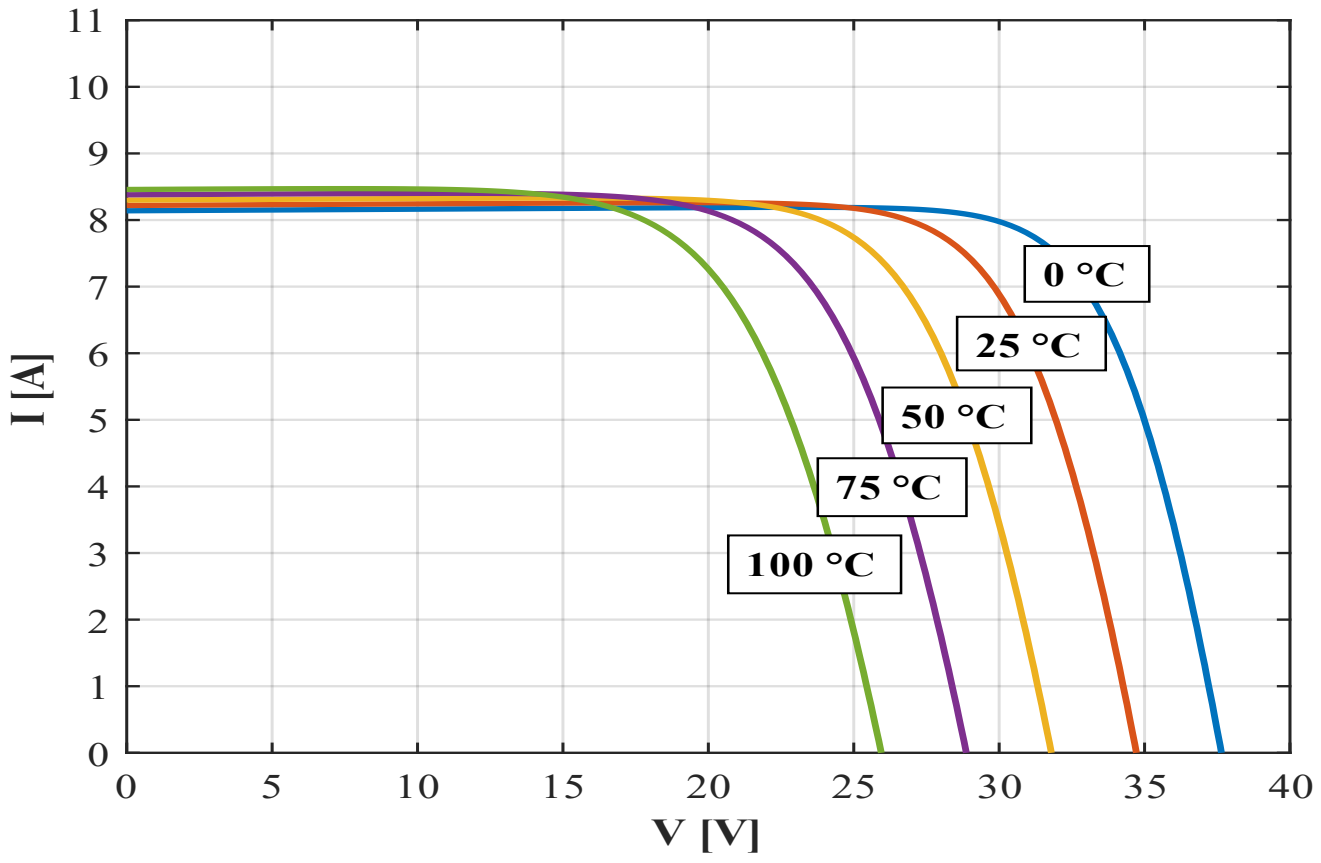

Fig. 2. I-V curves at different temperatures with fixed irradiance

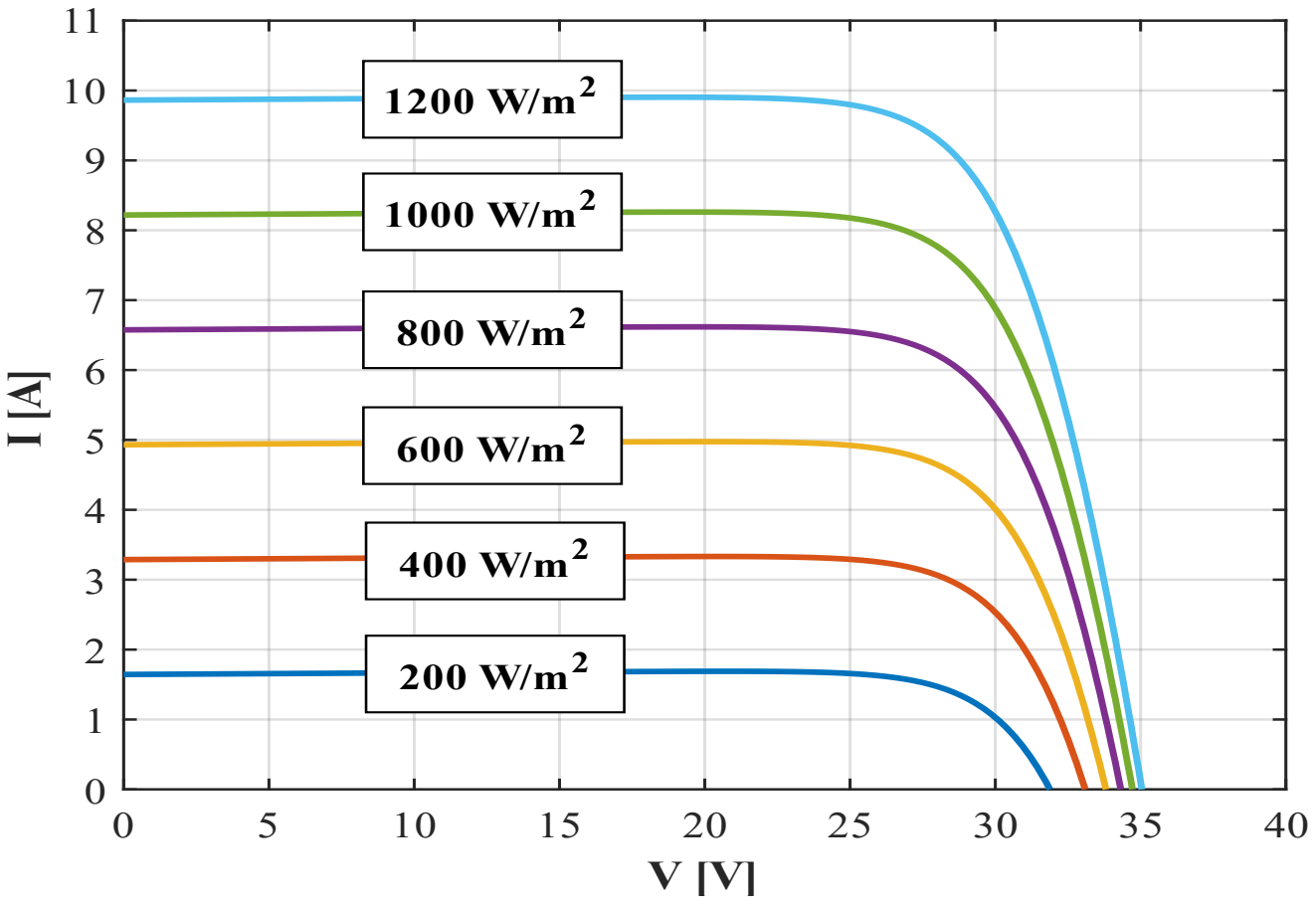

Fig. 3. I-V curves at different irradiance with fixed temperature

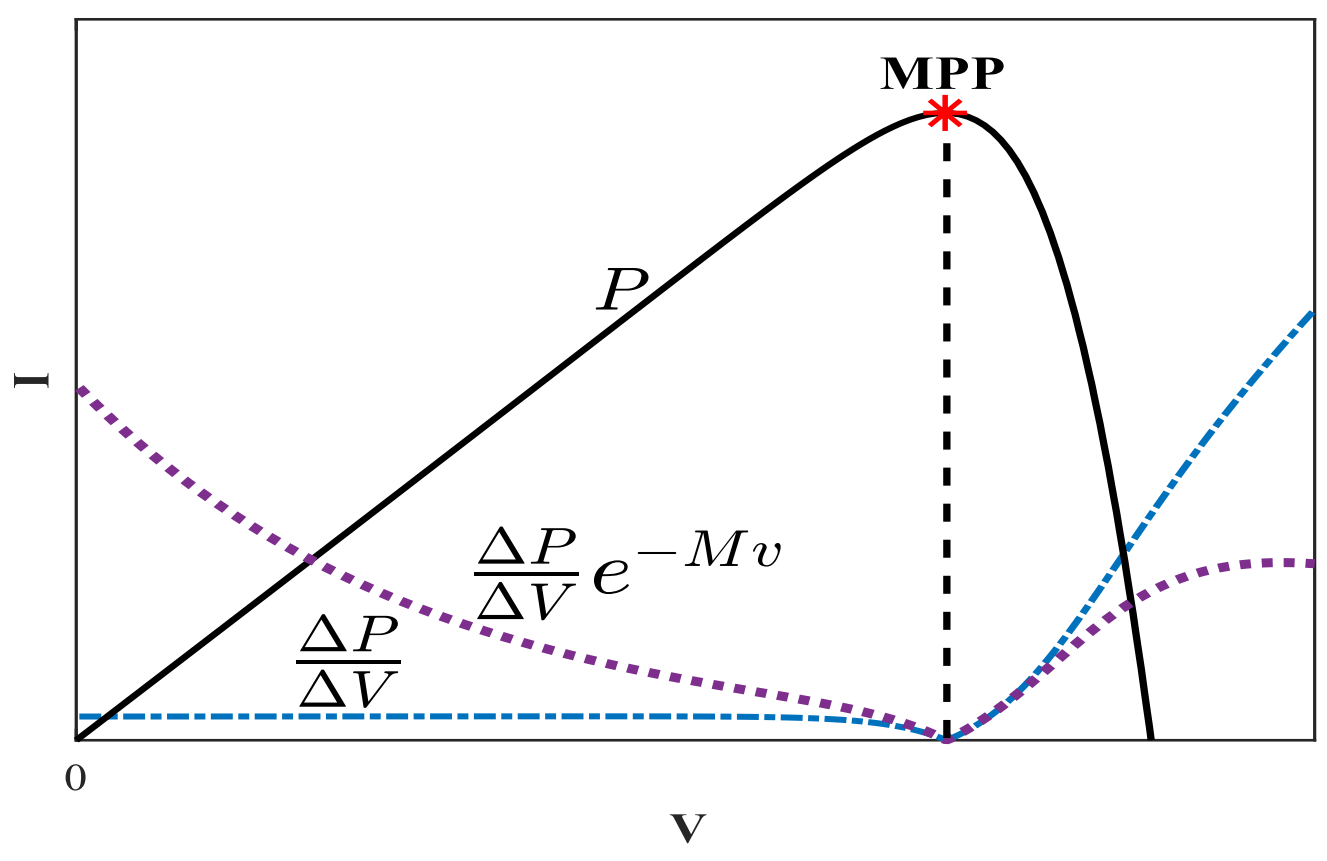

Fig. 4. P-V curve overlapped with ΔP/ΔV curves

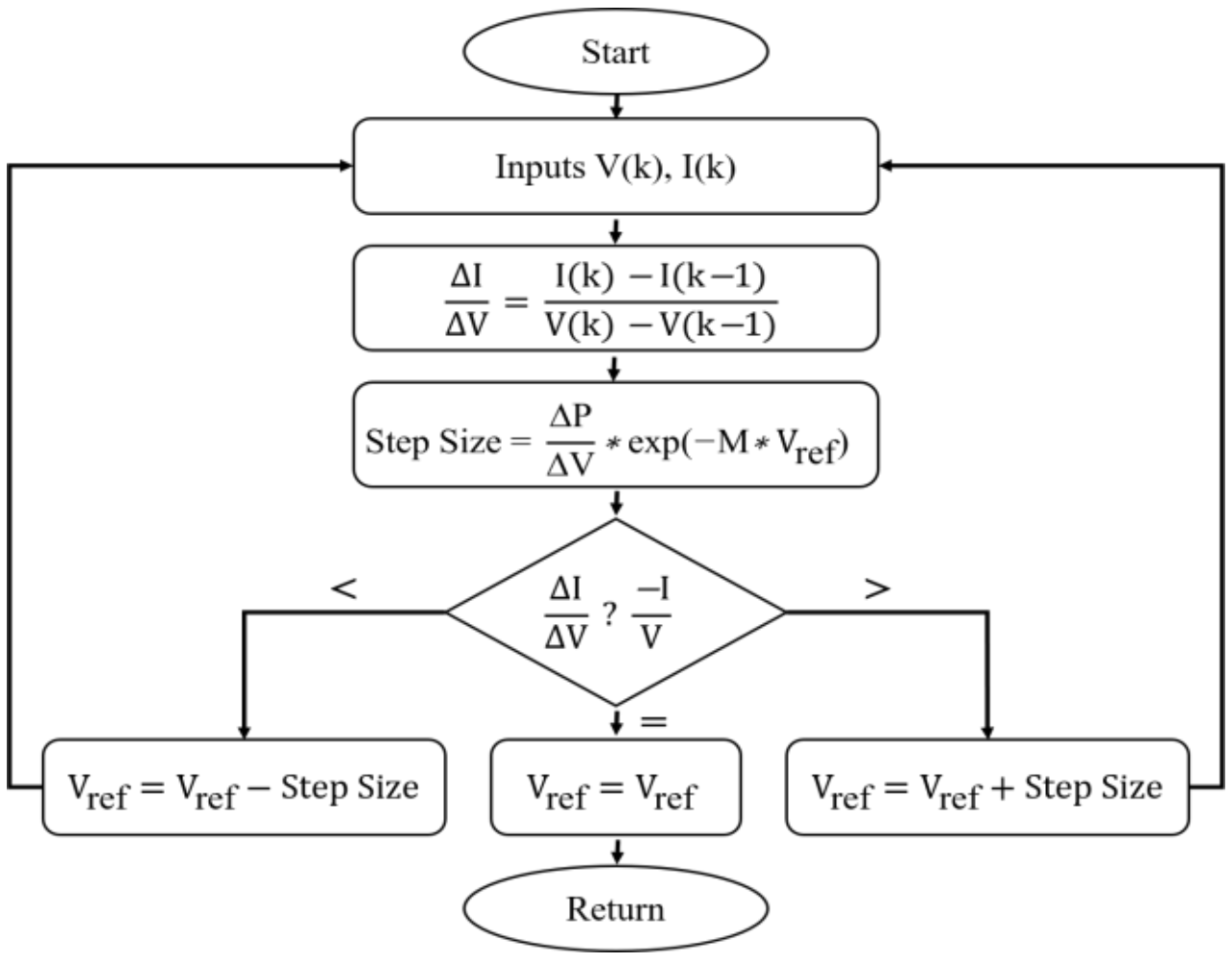

Fig. 5. Flow chart of the adaptive step size IC MPPT algorithm

Traditional IC technique increments (or decrements) the operating voltage by a fixed amount (fixed step size) until it reaches $V_{mp}$. This requires a smaller step size, which results in a large number of iterations if high accuracy is to be obtained. If a greater step size is used for faster convergence speed, the loss of accuracy is inevitable. This trade-off between convergence speed and accuracy can be mitigated with the IC technique with adaptive step size. In the adaptive step size IC technique, the step size is increased as the operating point gets further from the MPP and is decreased as the operating point approaches the MPP. This adaptive step size allows identification of the MPP with both high accuracy and fast convergence speed.

Different ways of choosing step size have been proposed previously. Hossain M. J. et al. proposed a method which uses a user defined exponential factor M and the incremental conductance in the feedback system in order to determine the necessary step size for fast convergence speed and high accuracy [7], while the algorithm proposed by Li. C. et al. does not require a user predefined constant and tracks the MPP through step size adjustment coefficients [8]. In this work, we propose a method which adjusts the step size based on the product of a decaying exponential function and ΔP/ΔV.

## II. Incremental Conductance Algorithm With Adaptive Step Size

As previously mentioned, ΔP/ΔV at the MPP must be zero and step size needs to be gradually decreased as the operating point approaches the MPP. This leads to the proposed method in which the step size is adjusted in terms of ΔP/ΔV. Fig. 4 shows the ΔP/ΔV-V curve along with the P-V curve, and it is shown that ΔP/ΔV in fact approaches zero as it approaches the MPP. Note that on the left side of the MPP, ΔP/ΔV stays nearly constant whereas it rapidly increases on the right side of MPP. This could result in instability and slow convergence speed in a rapidly changing environment. Therefore, as shown in Fig. 4, the exponential decaying function exp(-M*$v$) is multiplied in order to obtain the desired curve, where M is a predefined constant and $v$ is the operating voltage.

The detailed algorithm is described in the flowchart in Fig. 5. The incremental conductance (ΔI/ΔV) is first measured, then it is compared with the negative instantaneous conductance (-I/V). If ΔI/ΔV is less than -I/V, operating voltage is decremented by the step size. Similarly, if ΔI/ΔV is greater than -I/V, operating voltage is incremented by the step size. This process is then iterated until the MPP is found. At each iteration, the step size is updated to ΔP/ΔV *exp(-M*$v_{ref}$), which allows for the MPP to be determined with far fewer iterations compared to the fixed step size IC technique.

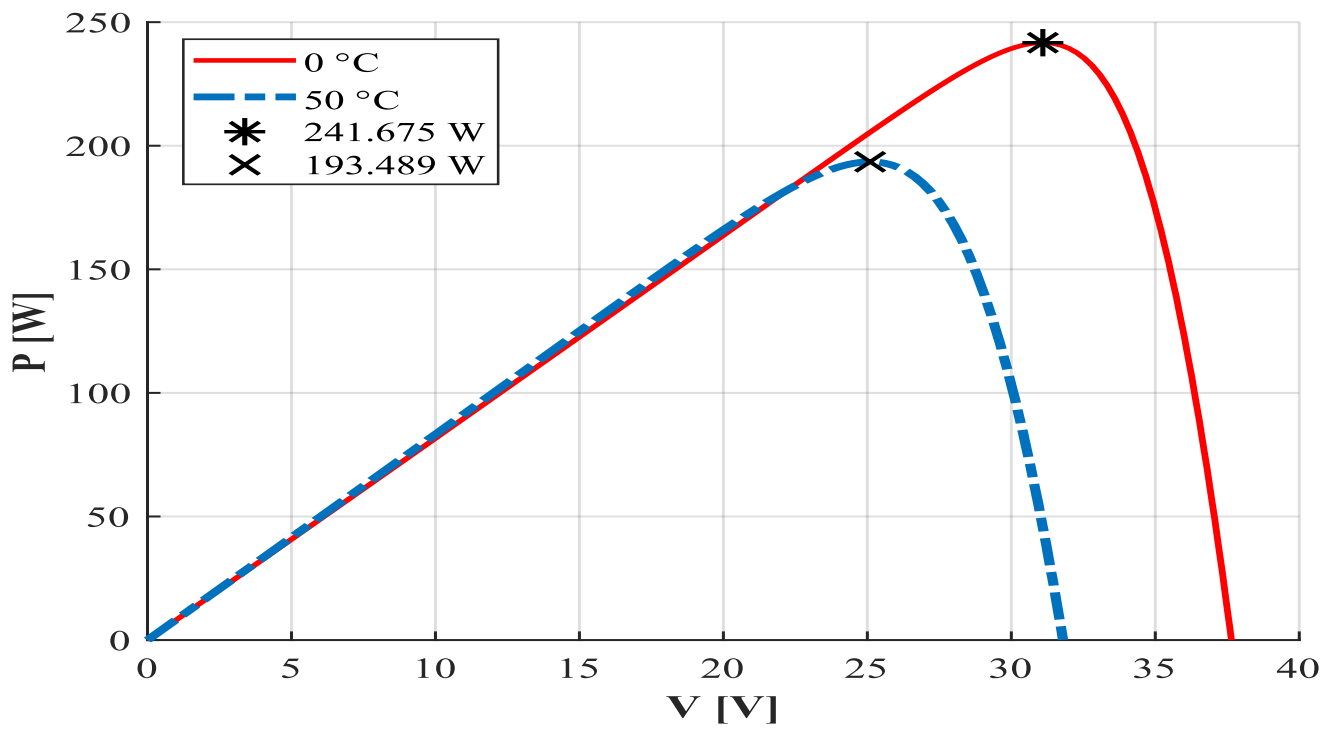

Fig. 6. Shift of MPP due to change in temperature (0°C to 50°C)

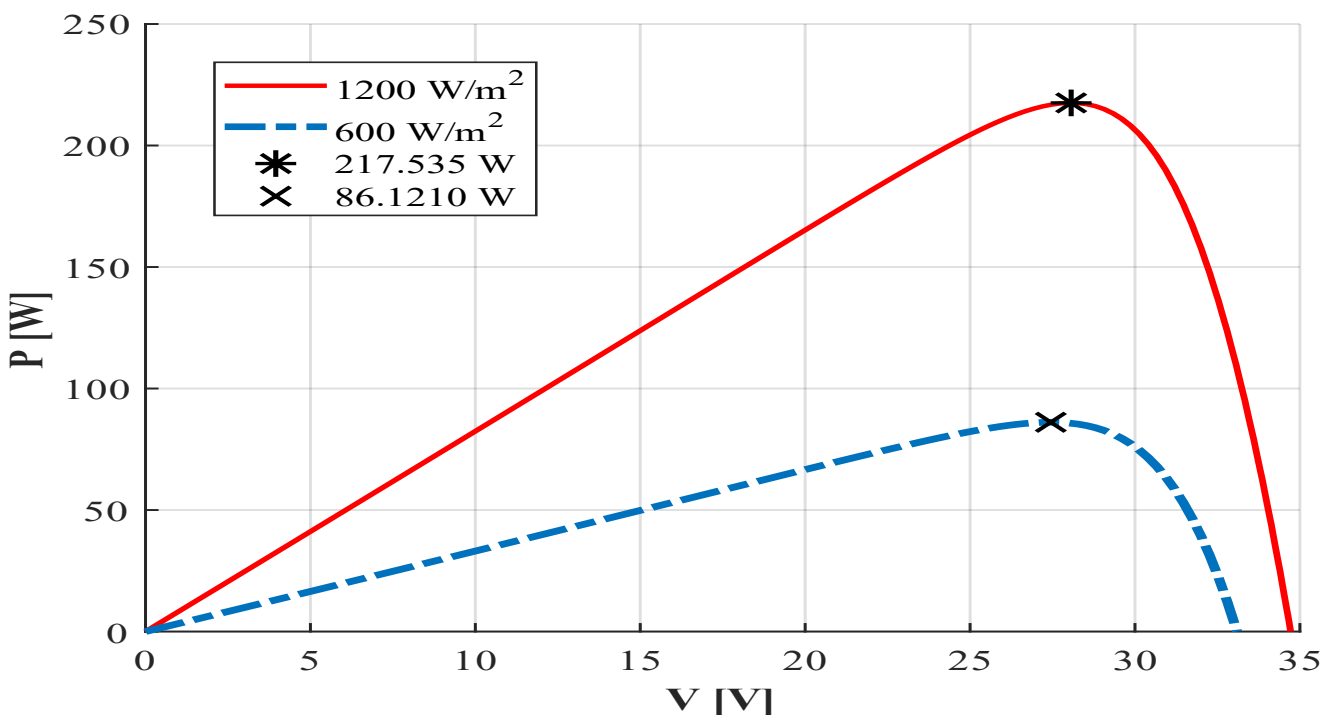

Fig. 8. Shift of MPP shift due to change in irradiance (600 W/m$^2$ to 1200 W/m$^2$)

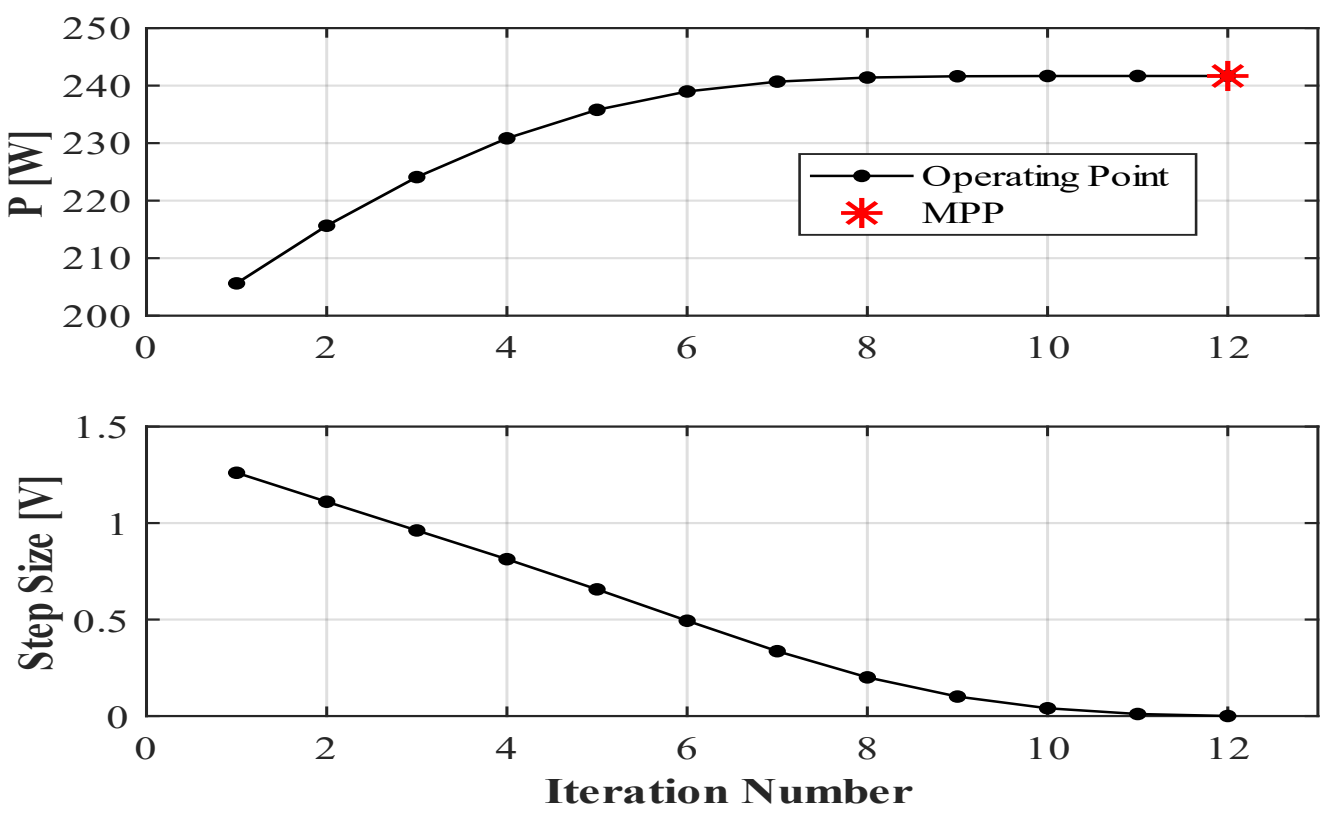

Fig. 7. Updated operating point and step size at each iteration

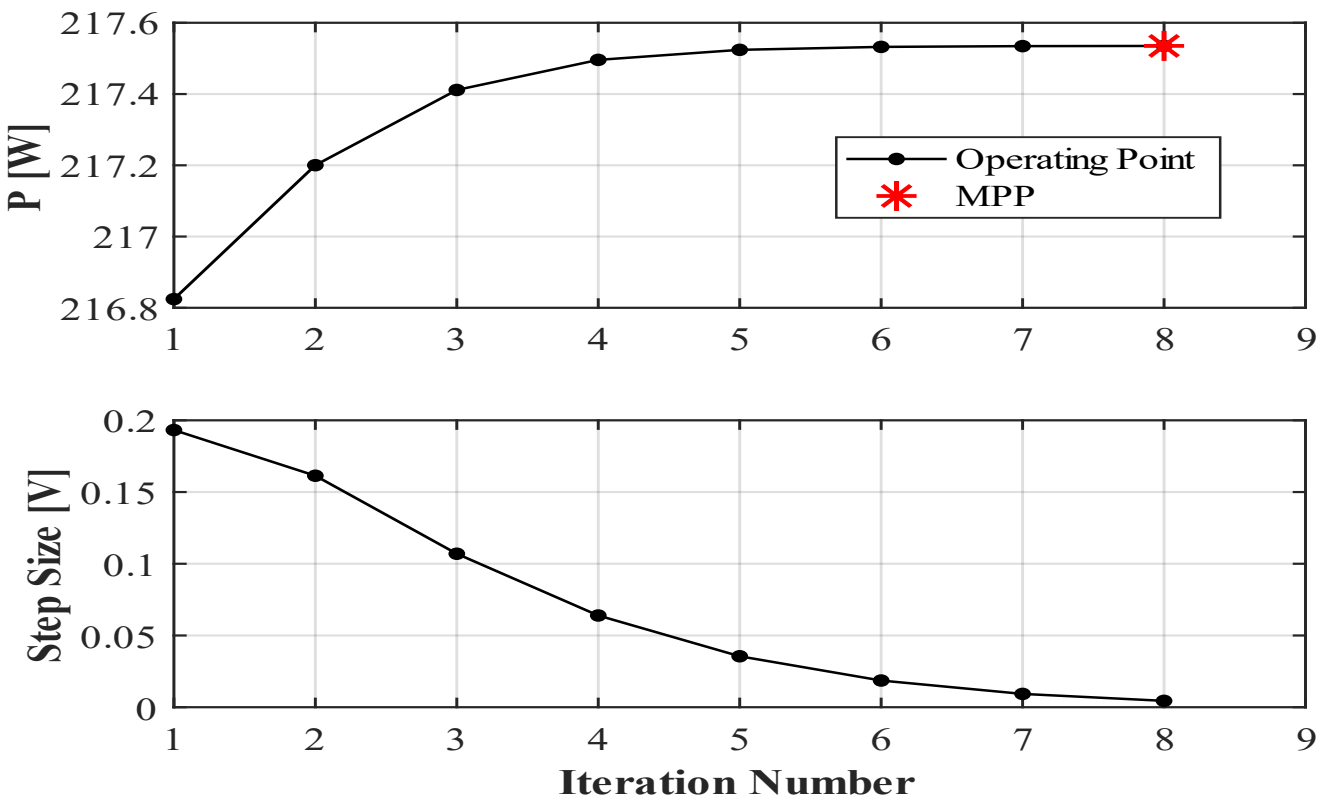

Fig. 9. Updated operating point and step size at each iteration

## III. Simulation Results and Anaalysis

A single-diode PV model and the parameters of the Kyocera KC200GT solar array were used for the simulation [1]. Simulations were done varying the temperature and irradiance, and using both the adaptive step size IC and the fixed step size IC techniques for comparison. The predefined constant M was chosen to be 0.09 for the adaptive step size IC technique. For the first simulation, irradiance was kept constant at 1000 W/m$^2$ and temperature was decreased from 50°C to 0°C. This results in shifting $V_{mp}$ from 25.0962 V to 31.0948 V and the maximum power from 193.489 W to 241.6751 W as shown in Fig. 6. Using the traditional IC technique with a fixed step size of 0.01, the maximum power was determined to be within the error of 0.00000213% of the actual maximum power and a total number of 599 iterations were recorded. Using the same fixed step IC technique, but with a fixed step size of 0.1 instead, the maximum power was determined to be 235.4256 W, which is within the error of 2.586% of the actual maximum power. However, the total number of recorded iterations was only 40. Here, the trade-off between convergence speed and accuracy of a fixed step size IC technique is clearly shown. Now, using the adaptive step size IC method, the maximum power was determined to be within the error of 0.0000359%, taking only 12 iterations. This result demonstrates that with the adaptive step size IC method, accuracy is significantly increased from 2.586% error to 0.0000359% error, and faster convergence speed is also achieved, going from 599 iterations down to only 12 iterations. Fig. 7 shows how the step size adaptively changes, approaching zero exponentially as the operating point approaches the MPP.

To investigate a scenario where the MPP shifts towards left, the temperature was increased from 0 °C to 50 °C with the irradiance kept constant at 1000 W/m$^2$. With the fixed step size IC technique, determined MPP was within the error of 0.000000635%, and it took 601 iterations to converge, whereas with the adaptive step size IC technique, it only took 20 iterations to find a MPP that is within the error of 0.0163%. Although the error is greater with adaptive step size, 0.0163% error results in only 0.0316 W of power loss. Furthermore, it takes 30 times fewer iterations to converge, 601 iterations to 20 iterations.

Finally, the impact of change in irradiance was also investigated. For this simulation, it was assumed that irradiance changed from 600 W/m$^2$ to 1200 W/m$^2$ with the temperature fixed at 25°C. Fig. 8 shows that $V_{mp}$ changes from 27.448 V to 28.064 V and the maximum power changes from 86.1210 W to 217.535 W. Therefore, the MPP shifted toward the right. Using the fixed step size IC technique, the determined MPP was within the error of 0.00000594%, taking 61 iterations. With the fixed step size IC technique, the maximum power was determined to be 217.53498 W, which is within the error of 0.0000620%, taking only 8 iterations. Because of the exponential nature of the changes in step size, it is noticed that the superiority of our adaptive step size IC technique compared to the fixed step size IC technique is more pronounced for the greater shift in $V_{mp}$.

In all of the above cases, adaptive step size IC technique permitted the MPP to be found with a much faster convergence speed and nearly the same level of negligible error as the fixed step size IC technique. More simulations were performed at different temperatures and irradiances to test the robustness of the algorithm, and the results are summarized in Table I-III.

## IV. Conclusion

An adaptive step size IC based MPPT algorithm has been proposed in this paper. This algorithm finds the MPP with increased convergence speed and nearly zero power loss compared to a fixed step size IC based MPPT algorithm. A series of simulations involving changes in temperature and irradiance were performed to test the robustness of the proposed algorithm, and it was shown that the proposed algorithm indeed surpasses the fixed step size algorithm at convergence speed while maintaining nearly the same level of accuracy.

TABLE I
SIMULATION RESULTS FROM VARYING TEMPERATURE WITH FIXED IRRADIANCE

| **Change in Temperature [°C]** | | **Change in Irradiance [W/m$^2$]** | | **Change in MPP [W]** | | **Accuracy [%]** | | **Number of Iterations Taken to Find MPP** | |
|---|---|---|---|---|---|---|---|---|---|
| Initial | Final | Initial | Final | Initial | Final | Fixed Step Size | Adaptive Step Size | Fixed Step Size | Adaptive Step Size |
| 25 | 0 | 1000 | 1000 | 217.54 | 241.68 | 1.327e-9 | 6.438e-5 | 302 | 17 |
| 50 | 15 | 1000 | 1000 | 193.49 | 227.18 | 8.029e-6 | 3.140e-5 | 416 | 16 |
| 12 | 30 | 1000 | 1000 | 230.08 | 212.72 | 4.749e-5 | 4.192e-4 | 217 | 23 |
| 5 | 60 | 1000 | 1000 | 236.84 | 183.91 | 4.468e-5 | 5.645e-3 | 656 | 38 |

TABLE II
SIMULATION RESULTS FROM VARYING IRRADIANCE WITH FIXED TEMPERATURE

| **Change in Temperature [°C]** | | **Change in Irradiance [W/m$^2$]** | | **Change in MPP [W]** | | **Accuracy [%]** | | **Number of Iterations Taken to Find MPP** | |
|---|---|---|---|---|---|---|---|---|---|
| Initial | Final | Initial | Final | Initial | Final | Fixed Step Size | Adaptive Step Size | Fixed Step Size | Adaptive Step Size |
| 0 | 0 | 600 | 900 | 145.16 | 217.71 | 3.080e-6 | 1.083e-4 | 15 | 8 |
| 0 | 0 | 1050 | 170 | 253.61 | 40.821 | 3.168e-3 | 6.591e-5 | 134 | 21 |
| 0 | 0 | 330 | 970 | 79.454 | 234.50 | 6.057e-5 | 5.594e-5 | 64 | 12 |
| 0 | 0 | 725 | 575 | 175.49 | 139.08 | 2.059e-4 | 3.301e-4 | 12 | 3 |

TABLE III
SIMULATION RESULTS FROM VARYING BOTH IRRADIANCE AND TEMPERATURE

| **Change in Temperature [°C]** | | **Change in Irradiance [W/m$^2$]** | | **Change in MPP [W]** | | **Accuracy [%]** | | **Number of Iterations Taken to Find MPP** | |
|---|---|---|---|---|---|---|---|---|---|
| Initial | Final | Initial | Final | Initial | Final | Fixed Step Size | Adaptive Step Size | Fixed Step Size | Adaptive Step Size |
| 45 | 0 | 200 | 900 | 38.058 | 217.71 | 2.235e-7 | 6.765e-5 | 701 | 25 |
| -5 | 20 | 525 | 725 | 161.17 | 129.57 | 3.662e-5 | 6.284e-4 | 292 | 33 |
| -12 | 7 | 1000 | 250 | 253.28 | 58.164 | 1.268e-3 | 1.054e-4 | 331 | 46 |
| 0 | 50 | 980 | 50 | 236.89 | 9.062 | 3.192e-2 | 1.527e-3 | 954 | 77 |